\documentclass[peer-review]{fa2026}

\title{Effects of HRTF Augmentation on Predicted Spatial Release from Masking in Music}

\author[1]{Jack Webb}
\author[2]{Christophe Lesimple}
\author[2]{Volker Kuehnel}
\author[1]{Lorenzo Picinali}
\correspondingauthor{j.webb24@imperial.ac.uk.}{Webb et al.}

\affil[1]{Dyson School of Design Engineering, Imperial College London, United Kingdom}
\affil[2]{Sonova AG, Stäfa, Switzerland}

\begin{document}
\maketitle

\begin{abstract}
Separating individual musical instruments within a complex mixture of sounds poses a persistent challenge for listeners with hearing loss. Although spatial separation of sources improves speech recognition in this population, the potential benefits of spatial cue enhancement for music perception remain largely unexplored. This paper introduces a method to increase spatial cue salience through the augmentation of individual head-related transfer functions (HRTFs). Auditory model analyses indicate that augmented HRTFs may enhance the separability of musical instruments relative to individual HRTFs. Predicted benefits persist when moderate sensorineural hearing loss is modelled, though they are substantially reduced. Simulated hearing aid processing does not restore these benefits to normal-hearing levels.
\end{abstract}

\section{Introduction}\label{sec:introduction}

Music listening for individuals with hearing loss is complicated by both the signal distortions introduced by hearing aids and a generally reduced acuity in music perception tasks \cite{greasley:26}. In particular, the perceptual separation of individual musical instruments in a mixture of sounds remains a significant hurdle \cite{greasley:20}. In essence this is a problem of auditory scene analysis, the process by which the auditory system decomposes complex sound environments into perceptually meaningful streams \cite{bregman:94}. The ability to perform auditory scene analysis is directly influenced by masking, whereby the threshold for hearing one sound is raised by the presence of a competing one. Masking can be reduced through several mechanisms. Energetic release arises from spectral and modulation differences between target and masker sounds \cite{culling:17}. Higher-level processes such as attention, memory, and structural processing can also provide informational release from masking \cite{kidd:17}. In complex acoustic scenarios, the spatial separation of target and masker provides cues that support both types of masking release \cite{ihlefeld:08}.

\subsection{Spatial Release from Masking in Music}

Spatial release from masking (SRM) provides sizeable advantages to speech intelligibility in situations with multiple overlapping talkers. Although more common in speech literature, we use SRM here to denote the general use of spatial cues to separate multiple auditory sources. An analogous task occurs in music scene analysis (MSA), defined as the ability to discern a single instrument from competing musical sounds. Studies have shown that MSA performance generally improves with fewer competing instruments and elevated target-to-masker level ratios \cite{hake:24}, as well as when fast dynamic range compression is used in hearing aids \cite{hake:25}, and when mixes have greater spectral domain sparsity \cite{benjamin:25}. However, the role of SRM in MSA remains elusive. Only one work has analysed the effect of spatial cue manipulations on MSA performance: \cite{hake:24} identified a small but inconsistent benefit of increased stereo width on MSA, induced by frequency-independent level differences between left and right headphone channels. As such, the role of more complex spatial auditory cues in facilitating the segregation of music sources remains unclear.

A natural extension is to consider more complete spatial representations based on head-related transfer functions (HRTFs). Spatial auditory perception relies on interaural differences in time (ITD) and level (ILD), as well as monaural spectral cues. HRTFs encapsulate these cues, characterising how sound is directionally filtered by an individual listener's morphology before it reaches the eardrum. Unlike intensity stereo panning, which relies solely on frequency-independent ILDs, binaural rendering using HRTFs provides the auditory system with all three types of spatial cue, motivating exploration of their role in MSA.

\subsection{Augmentation of Spatial Cues}

Speech intelligibility research has investigated the possibility of increasing SRM by enhancing spatial cues. One approach involves designing filters that transform binaural signals to achieve linear magnification of interaural time and level differences \cite{durlach:86}. Such algorithms have been shown to improve speech intelligibility in normal-hearing listeners \cite{kollmeier:90}. Low frequency and broadband ILD magnification have also improved speech intelligibility in bilateral cochlear implant users \cite{richardson:25}. Other works have achieved similar magnification outcomes by augmenting the interaural cues contained within HRTFs; in \cite{marggraf:25}, augmentation of individually measured HRTFs resulted in improved speech intelligibility. Augmented HRTF spectra have also produced improvements in vertical localisation accuracy \cite{brungart:09}. Despite potential benefits, all of these approaches have yet to be evaluated in MSA contexts.

\subsection{Effects of Hearing Loss}

Sensorineural hearing loss is known to contribute to compromised spatial auditory processing, leading to generally reduced SRM benefit \cite{arbogast:05}. Such deficits likely disrupt auditory scene analysis in listeners with hearing loss, whether this manifests as increased thresholds to understand speech in noise \cite{goossens:17}, or increased thresholds to hear specific instruments within a mixture \cite{siedenburg:20}. Understanding the potential effects of hearing loss is essential to evaluating the utility of spatial cue augmentation for music listening in assistive devices.

\subsection{Aims}
Given the limited understanding of SRM in MSA, the uncertain impact of augmented spatial cues on SRM in music listening, and the potential deficits that may be introduced by hearing loss, the aims of this study are threefold:

\begin{enumerate}
    \item To outline a framework for systematically augmenting interaural cues within individual HRTFs.
    \item To estimate, using computational auditory models, the relative contribution of natural and augmented spatial cues to SRM in music listening.
    \item To assess the likely influence of sensorineural hearing loss on access to SRM in music.
\end{enumerate}

As this study relies on computational models of SRM and hearing loss, results are interpreted as predictions that provide a basis for future behavioural validation.

\section{Methods}
\subsection{HRTF Augmentation}
In a departure from previous augmentation approaches, we propose to enhance spatial cues by scaling the direction-dependent spectral components of a HRTF. While principal component analysis (PCA) has been long used in HRTF reconstruction \cite{kistler:92}, its application to spatial cue augmentation remains mostly unexplored. We performed PCA on HRTFs drawn from 405 individuals within the SONICOM HRTF dataset \cite{poole:25}.

Let $h$ represent the time-domain head-related impulse response (HRIR) from an arbitrary source position, with corresponding HRTF $H$ in the frequency domain. Firstly, the directional transfer function (DTF) is obtained from each HRTF by subtracting the common transfer function (CTF) - the mean log-magnitude response averaged across all spatial locations - from each HRTF. The DTF is then decomposed into mid and side components. For left and right ear DTF spectra $L[k]$ and $R[k]$,

\[
M[k] = \tfrac{1}{2}\bigl(L[k] + R[k]\bigr), \qquad
S[k] = \tfrac{1}{2}\bigl(L[k] - R[k]\bigr),
\]

where $S$ captures interaural spectral differences including frequency-dependent ILDs. PCA is performed on this side channel $S$ over all horizontal plane source positions and listeners. Observations were stacked into a matrix and decomposed by singular value decomposition as $S = U\Sigma V^\top$, where $V$ contains the spectral basis functions and $s_i = U_i \Sigma$ gives the PCA scores for observation $i$. To magnify direction-dependent interaural structure, we apply residual contrast in score space: for each listener, we compute the mean side-channel score vector $\boldsymbol{\mu}$, averaged across horizontal directions, and scale only the deviation from this mean

\[
\hat{\mathbf{s}}_i = \alpha \bigl(\mathbf{s}_i - \boldsymbol{\mu}\bigr),
\]
using the first six principal components. All other scores are set to zero. The augmented side spectrum is reconstructed as $\hat{S}_i[k] = \hat{\mathbf{s}}_i V^\top$, and the augmented DTFs are rebuilt from the unchanged $M$ component:

\[
\hat{L}_i[k] = M_i[k] + \hat{S}_i[k], \qquad
\hat{R}_i[k] = M_i[k] - \hat{S}_i[k].
\]

\begin{figure}[ht]
\centering
 \includegraphics[width=7cm]{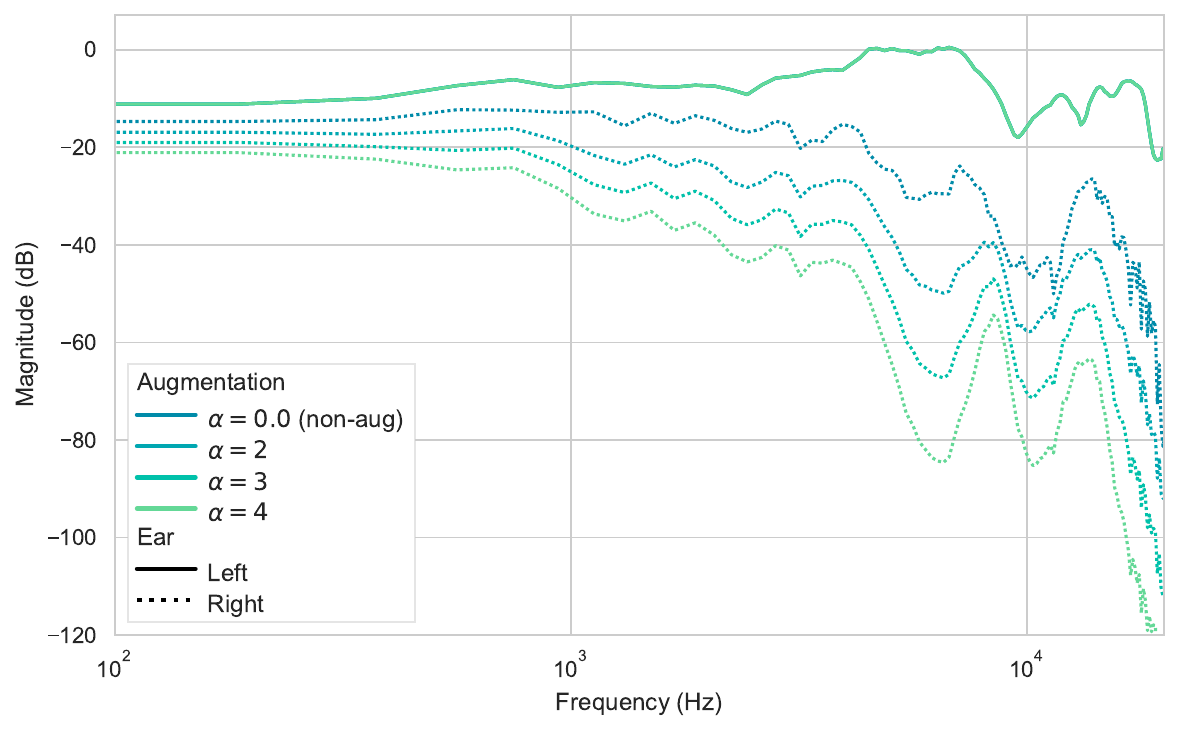}
 \caption{Example individual and augmented HRTFs for a source at $90^\circ$ azimuth on the horizontal plane.}
 \label{fig:hrtfs}
\end{figure}

The listener-specific CTF is then added back to obtain the augmented log-magnitude HRTF. The final augmented HRTF $\widehat{H}_i$ is obtained by exponentiating the augmented magnitude and restoring the original phase ($\angle H_i$). To minimise colouration, only the contralateral ear is augmented in each HRTF, and midline source positions are left unaltered. This formulation magnifies frequency-dependent ILDs while preserving broader monaural structure (Fig.~\ref{fig:hrtfs}).

\subsection{Spatial Release from Masking Analysis}

\subsubsection{Stimuli}

Numerical analysis was conducted to estimate the SRM benefit of different spatial cue combinations. A large dataset of musical samples was constructed using multitracks drawn from the Musiclarity and MedleyDB datasets \cite{eastgate:16, bittner:16}. Each sample was 2 seconds in duration, and consisted of a single target instrument plus three masking instruments. The samples were each processed through seven rendering conditions: \textit{Diotic}, fully diotic mix; \textit{ILD}, frequency-independent ILDs matched to the broadband RMS ILD of each HRTF; $\textit{ILD+ITD}$, frequency-independent ILDs plus onset-threshold-estimated ITDs matched to each HRTF; \textit{HRTF}, individually measured HRTFs; and $\textit{HRTF}_{2.0}$, $\textit{HRTF}_{3.0}$, and $\textit{HRTF}_{4.0}$, augmented HRTFs with $\alpha=2.0$, $3.0$, and $4.0$, respectively.

All sources were restricted to the horizontal plane. Target instruments were evenly divided across four categories (guitar, piano, synthesiser, and drums). Two geometries were tested: target at $0\text{\textdegree}$ azimuth with maskers divided between $\pm60\text{\textdegree}$, and target at $\pm60\text{\textdegree}$ with one masker at $0\text{\textdegree}$ and the remaining maskers at the opposite $60\text{\textdegree}$ location. Eight samples were used per target category, each rendered under the seven rendering conditions and both spatial geometries. This was repeated 25 times, with each repeat using a unique HRTF from the SONICOM HRTF dataset corresponding to one individual listener \cite{poole:25}. Sample selection was randomised independently for each listener profile. For all stimuli, the target-to-masker level ratio was fixed at $-6$ dB RMS.

\subsubsection{Auditory Model Analysis}

To predict spatial release from masking, we employed two models available in the Auditory Modeling Toolbox \cite{majdak:22}. First, SRM was estimated using the \textit{vicente2020} binaural speech intelligibility model \cite{vicente:20}. Second, we applied the \textit{bischof2023} model, originally designed to predict the unmasking of reverberant complex tones in noise \cite{bischof:23}. Both \textit{vicente2020} and \textit{bischof2023} combine better-ear SNR (BE SNR) and binaural unmasking (BU) to estimate effective SNR or total unmasking, respectively. Unlike \textit{vicente2020}, \textit{bischof2023} does not incorporate speech intelligibility weighting across frequency bands, providing a comparison that isolates the effect of speech weighting. It also utilises $12$\,ms analysis windows that allow it to implicitly account for the short-term monaural glimpsing effects explicitly modelled by \textit{vicente2020}, making it better suited to non-stationary music signals than SRM models that assume relatively static inputs. SRM was defined as the dB benefit of spatial cues relative to the diotic condition, computed by subtracting the diotic score from the corresponding spatial condition score for each sample. Results were compared across conditions and target instrument categories.

\subsubsection{Hearing Loss Simulation}

Analysis with \textit{vicente2020} was conducted using three hearing profiles: an N0 audiogram representing normal hearing, an N3 audiogram representing a moderate sensorineural hearing loss profile, and an N3 audiogram with samples pre-processed through a wide dynamic range compression (WDRC) hearing aid simulator. Hearing loss was modelled with \textit{vicente2020}, which uses internal noise to capture threshold elevation and level-dependent coding deficits. Target and masking instruments were scaled together so that the broadband level of the maskers was fixed at $65$\,dB SPL prior to any amplification.

\section{Results}

\subsection{SRM with Normal Hearing}

\begin{figure}[h]
\centering
 \includegraphics[width=7.6cm]{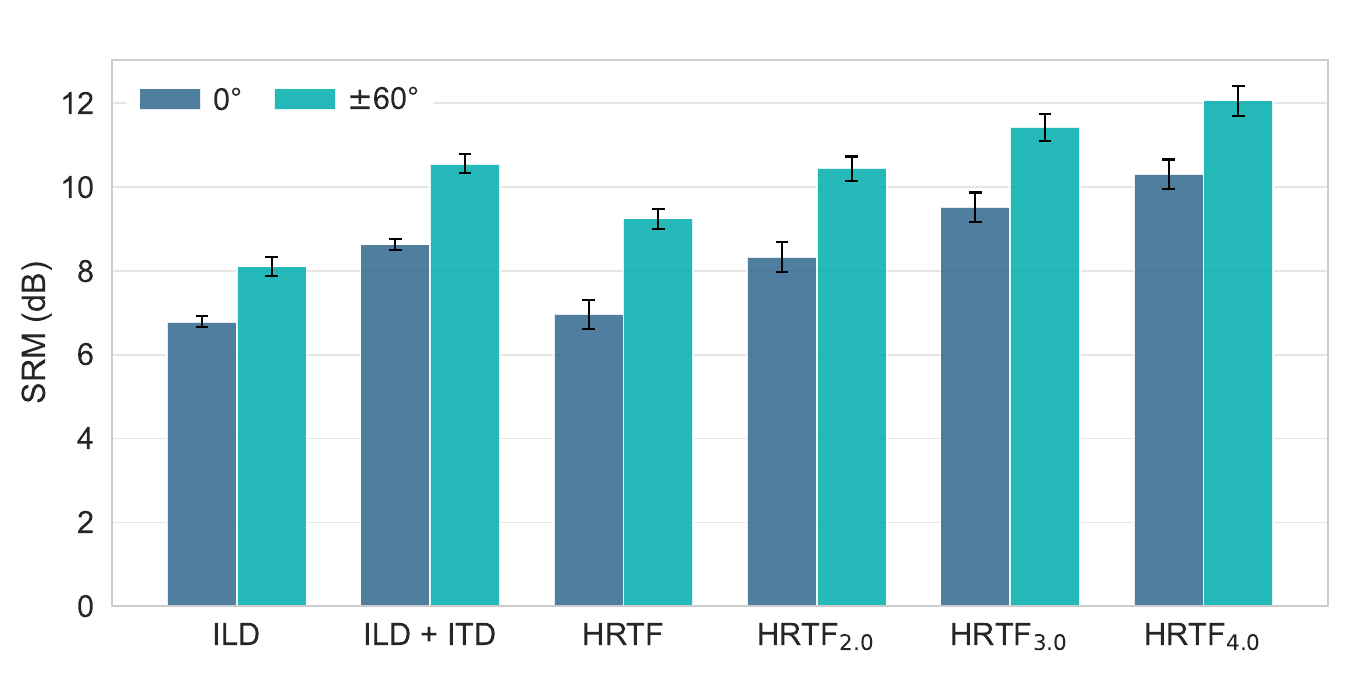}
 \caption{Mean predicted SRM relative to diotic baseline, split by target azimuth. Higher SRM indicates increased potential to discern the target instrument. Results are from \textit{vicente2020} model with an N0 profile; error bars indicate bootstrapped 95\% confidence intervals from 10,000 samples clustered by listener profile.}
 \label{fig:srm}
\end{figure}

Fig. \ref{fig:srm} displays the mean SRM benefit predicted by \textit{vicente2020} with an N0 audiogram, averaged across target instrument categories. All conditions elicit an SRM benefit relative to the diotic baseline. With a frontal target, individual HRTFs were predicted to provide $6.97$ dB SRM, only a marginal increase of $0.18$ dB over the ILD condition. Significantly, predicted SRM increased with the magnitude of HRTF augmentation: $\text{HRTF}_\text{{4.0}}$ showed the largest predicted SRM of $10.31$ dB for a frontal target. Across conditions, SRM was higher when the target was located at $\pm60\text{\textdegree}$ than at $0\text{\textdegree}$, on average by $1.88$ dB.

For the augmented HRTF conditions, increases in SRM were mainly driven by improvements in BE SNR, with only minor reductions in BU. From individual HRTFs to $\text{HRTF}_\text{{4.0}}$, mean N0 SRM across target locations increased by $3.08$ dB, comprising a $3.35$ dB increase in BE SNR and a $0.27$ dB reduction in BU. A similar pattern was observed under \textit{bischof2023}, except that the ILD+ITD condition exceeded $\text{HRTF}_\text{{3.0}}$ in terms of SRM. Relative to the ILD condition, adding ITD increased SRM by 59\% under \textit{bischof2023} versus 29\% under \textit{vicente2020}.

\subsection{SRM with Hearing Loss}

\begin{figure}[ht]
 \includegraphics[width=7.8cm]{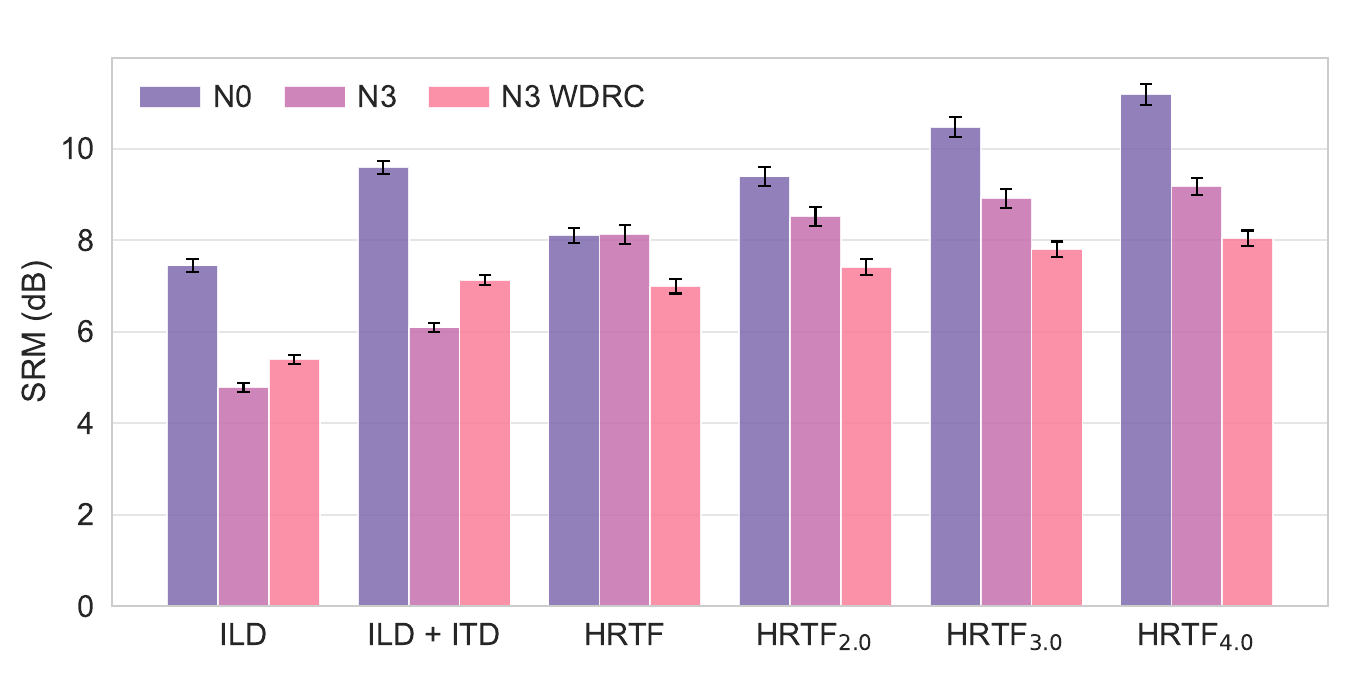}
 \centering
 \caption{Mean SRM predicted by \textit{vicente2020} averaged over target locations, split by hearing profile: normal hearing (N0), unaided hearing loss (N3), aided hearing loss (N3 WDRC). Error bars show bootstrapped 95\% confidence intervals.}
 \label{fig:hearing-loss}
\end{figure}

\begin{figure}[ht]
 \includegraphics[width=\linewidth]{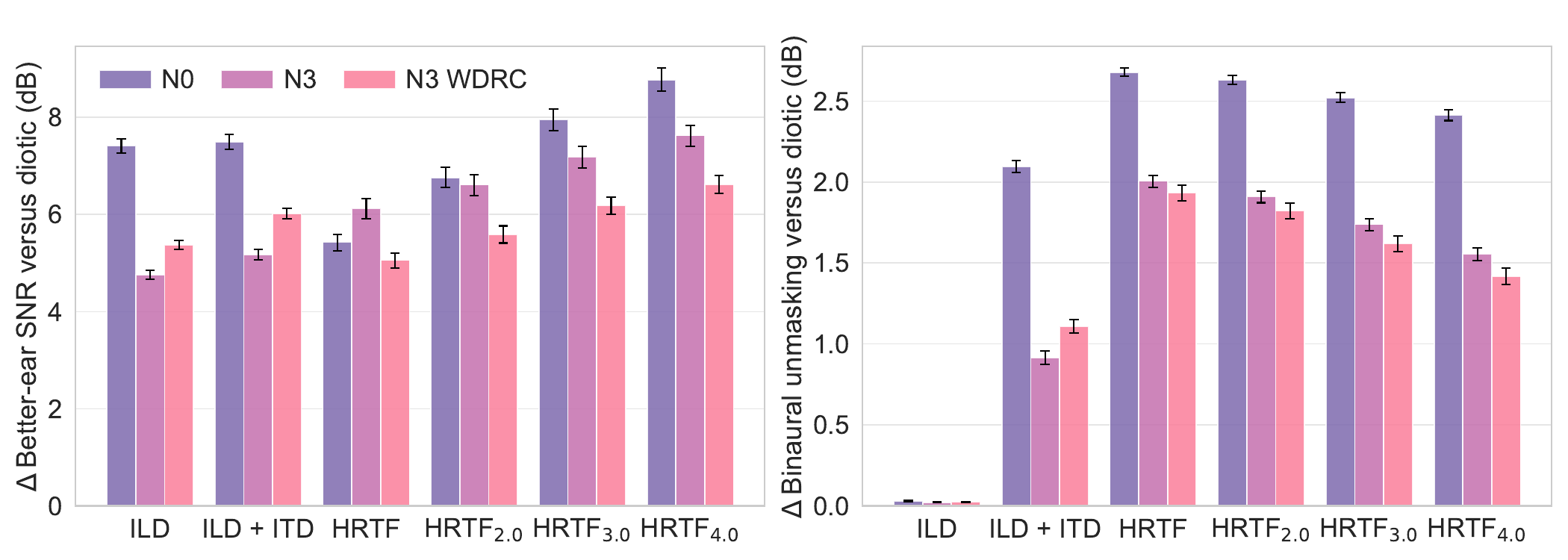}
 \caption{Mean change in BE SNR (left) and BU (right) relative to diotic condition, under \textit{vicente2020}. Results averaged over target locations and split by simulated hearing profile. Error bars show bootstrapped 95\% confidence intervals.}
 \label{fig:hearing-loss-be}
\end{figure}

As indicated in Fig. \ref{fig:hearing-loss}, modelling an N3 loss with \textit{vicente2020} resulted in a fall in mean SRM across conditions from $9.37$ to $7.60$ dB. Fig. \ref{fig:hearing-loss-be} divides the predicted SRM into BE SNR and BU contributions, each relative to the diotic condition. This reveals, across conditions, an average $0.65$ dB reduction in access to BU with an unaided N3 loss. Furthermore, BE SNR gains provided by HRTF augmentation were diminished: between individual HRTFs and $\text{HRTF}_\text{{4.0}}$, mean BE SNR gains fell from $3.35$ dB to $1.49$ dB for N0 and unaided N3 respectively, as access to higher frequency ILDs was lost. Though a significant SRM benefit remained for $\text{HRTF}_\text{{4.0}}$ versus the individual HRTFs with an N3 loss ($p < 0.001$), the magnitude of this gain decreased from $3.08$ dB to $1.05$ dB. 

Pre-processing samples through WDRC to simulate a hearing aid partially restored audibility for the N3 profile, raising \textit{vicente2020} mean effective SNR across conditions from $-6.84$ to $-2.76$ dB, but did not have a systematic effect on SRM across conditions. The SRM gain between $\text{HRTF}_\text{{4.0}}$ and individual HRTFs remained unchanged from the unaided N3 result ($\Delta = +0.01$ dB, $p = 0.81$). On average across all dichotic mixes, WDRC processing also resulted in $2.87$ dB distortion to short-term ($24$\, ms) broadband ILD. Larger ILD distortion correlated with reductions in SRM relative to N0 ($r = 0.31$, $p < 0.001$).

\subsection{Effect of Target Instrument}

Under N0, the SRM gain provided by $\text{HRTF}_\text{{4.0}}$ relative to individual HRTFs differed across target instruments (Friedman $p = 0.003$), with the synthesiser targets showing significantly higher SRM gain than drum targets ($3.39$ dB versus $2.80$ dB, Holm-adjusted $p = 0.007$). In the N3 and N3 aided conditions, no significant differences emerged between instruments.

\section{Discussion}

\subsection{Spatial Cue Contributions to MSA}

A data-driven augmentation method was outlined, in which principal components derived from a population of measured HRTFs were scaled to enhance frequency-dependent ILDs in individual HRTFs. For an N0 listener, auditory models predicted greater SRM with augmented than with individual HRTFs, suggesting that augmentation may support clearer instrument separation than is achievable using natural spatial cues. While augmentation benefits were largely attributed to elevated BE SNR, SRM predictions for an N0 listener also increased with the introduction of ITDs: predicted SRM was greater under the ILD+ITD condition versus the ILD condition. This deviates from the behavioural results of \cite{glyde:13}, who reported that ITDs added marginal SRM benefit in a speech-on-speech task when ILDs were already present. This discrepancy may be tied to high informational masking creating performance limits in that study, which the present approach does not account for. However, the significance of ITD-based equalisation-cancellation mechanisms in music remains an open question; unlike BU speech models designed to cancel competing talkers, MSA may prioritise the ability to hear multiple instruments simultaneously.

Across results, predicted SRM was influenced by the frequency distribution of spatial cues. For instance, \textit{bischof2023} predicted a larger relative benefit for the ILD+ITD condition than \textit{vicente2020}, plausibly because speech-weighting curves down-weight low-frequency SRM. Since music has more variable long-term spectra than speech \cite{chasin:04}, low frequencies may contribute more to SRM than speech-weighting functions assume, making a corresponding behavioural result likely. Additionally, for N0 listeners, \textit{vicente2020} predicted comparable SRM for the ILD and HRTF conditions, even though HRTFs also convey ITDs. This arises as the ILD condition was matched to the broadband ($80$Hz-$16$kHz) ILD of each individual HRTF, whereas ILDs in a HRTF vary with frequency. As such, target instrument spectral centroids fell at frequencies where ILDs in the HRTF condition were smaller than those in the flat ILD condition, on average by $5.09$ dB at $\pm60\text{\textdegree}$. Together, these results suggest that spatial cue benefits for MSA rely on how cue magnitude aligns with the target spectrum.

\subsection{Audibility Limits SRM Benefits}

With an unaided N3 loss, the SRM benefit of HRTF augmentation persisted but was largely diminished as elevated internal noise rendered higher frequency ILDs inaudible. Individual HRTFs were also predicted to yield greater SRM than broadband-matched ILD+ITD cues, reversing the N0 ordering. Further analysis suggested this was driven by an audibility effect. Under N3, elevated internal noise around $2$-$4$\,kHz 
shifted the primary determinant of SRM from head-shadowing advantages to pure target audibility. HRTF spectral colouration increased target energy in this region, allowing brief glimpses above the internal noise floor, whereas ILD+ITD provided no comparable spectrally local improvement. Hearing aid processing partially restored audibility in this region, thereby reinstating the N0 condition ordering. Overall though, average SRM with WDRC remained below normal-hearing levels, likely due to residual hearing loss effects and the interaural cue distortions introduced by unlinked bilateral WDRC processing.

\subsection{Limitations}

There are notable limitations to the presented modelling approach. First, SRM in music listening may be modulated by additional auditory scene analysis cues such as temporal modulation, harmonic structure, and pitch differences, in ways not captured by models of speech and complex tone unmasking. The hearing loss modelling is also approximate, and does not explicitly account for auditory filter broadening and reduced temporal fine structure processing \cite{vicente:20}. These omissions may fail to capture cues that are relevant for SRM in music listening. In particular, the lack of substantial differences in SRM gain between target instruments observed here, especially under an N3 audiogram, seems unlikely to be replicated behaviourally. Enhanced low-frequency ILDs have improved localisation in listeners with hearing loss only for stimuli with distinct envelope fluctuations \cite{moore:16}, suggesting that instrument-dependent temporal-envelope characteristics could influence behavioural results.

Second, objective models neglect higher-level processes. Given evidence that the auditory scene analysis of music and speech share perceptual mechanisms \cite{hake:25b}, behavioural SRM may diverge from model predictions as structural cues and listener adaptation modulate informational masking. For instance, \cite{gonzalez:24} observed significantly greater median-plane SRM with individual HRTFs than mannequin HRTFs. \cite{detaillez:18} found no speech intelligibility benefit from interaural magnification, partly attributing this to distorted spatial cues introduced by the magnification algorithm. Such unfamiliar cues may be harder to process, particularly given potential relationships between localisation accuracy and SRM performance \cite{srinivasan:16}. Augmented HRTFs may similarly depart from a listener’s learned spatial mapping, potentially explaining why \cite{marggraf:25} observed smaller behavioural SRM gains than predicted by auditory models. Nevertheless, since HRTF augmentation improved speech intelligibility in that study, we expect the present approach to yield benefits in behavioural music scene analysis tests.

\section{Conclusion}

A framework was introduced by which spectral cues embedded in HRTFs were augmented to increase the spatial release from masking of target instruments in music. Auditory model predictions indicated a benefit to the approach, with music rendered using augmented HRTFs achieving elevated SRM compared with individual HRTFs. This advantage persisted under simulation of moderate sensorineural hearing loss but was substantially reduced. Hearing aid processing partially restored overall audibility, but did not meaningfully recover augmentation-related SRM gains.

\section{Acknowledgments}
This work was supported by an EPSRC CASE conversion PhD scholarship, co-funded by Sonova. Our thanks to Michel Bürgel and Kai Siedenburg for their provision of the stimuli from \cite{hake:24}.

\printbibliography

\end{document}